\documentclass{elsart}

\usepackage{graphicx,epsfig}    

\newcommand{\heeg} {\mbox{$\et\rightarrow e^+e^-\gamma$}}
\newcommand{\hppee}{\mbox{$\et\rightarrow\pi^+\pi^-e^+e^-$}}
\newcommand{\hppg}{\mbox{$\et\rightarrow \pi^+\pi^-\gamma$}}
\newcommand{\hppp}{\mbox{$\et\rightarrow \pi^+\pi^-\pi^0$}}

\newcommand{\et}{\mbox{$\eta$}}
\newcommand{\pdHeh}{\mbox{$pd\to^3$He\,$\eta$}}
\newcommand{\pim}{\mbox{$\pi^-$}}
\newcommand{\pip}{\mbox{$\pi^+$}}

\newcommand{\E}[1]{\mbox{$\times$10$^{#1}$}}
\newcommand{\eefakeg}{\mbox{$e^+e^-$''$\gamma$''}}
\newcommand{\fakeg}{\mbox{''$\gamma$''}}

\begin{document}
\begin{frontmatter}

\runauthor{Chr.~Bargholtz et al.}

\title{Measurement of the \hppee\ decay branching ratio }
\collab{CELSIUS-WASA Collaboration}
\author[SU]{Chr.~Bargholtz},
\author[Tuebingen]{M.~Bashkanov},
\author[Dubna]{D.~Bogoslawsky},
\author[TSL]{H.~Cal\'en},
\author[UU]{F.~Cappellaro},
\author[Tuebingen]{H.~Clement},
\author[Hamburg]{L.~Demir\"ors}, 
\author[TSL]{C.~Ekstr\"om},
\author[TSL]{K.~Fransson},
\author[SU]{L.~Ger\'en},
\author[UU]{L.~Gustafsson},
\author[UU]{B.~H\"oistad},
\author[Dubna]{G.~Ivanov},
\author[UU]{M.~Jacewicz\thanksref{cor}},
\thanks[cor]{Corresponding author. {\sl E-mail address:} Marek.Jacewicz@tsl.uu.se}
\author[Dubna]{E.~Jiganov},
\author[UU]{T.~Johansson}, 
\author[UU]{S.~Keleta},
\author[UU]{I.~Koch},
\author[UU]{S.~Kullander},
\author[TSL]{A.~Kup\'s\'c},
\author[Dubna]{A.~Kuznetsov},
\author[MEPI]{I.V.~Laukhin},
\author[SU]{K.~Lindberg},
\author[TSL]{P.~Marciniewski},
\author[Tuebingen]{R.~Meier},
\author[Dubna]{B.~Morosov},
\author[Juelich]{W.~Oelert},
\author[Hamburg]{C.~Pauly},
\author[UU]{H.~Pettersson},
\author[Dubna]{Y.~Petukhov},
\author[Dubna]{A.~Povtorejko},
\author[TSL]{R.J.M.Y.~Ruber},
\author[UU]{K.~Sch\"onning},
\author[Hamburg]{W.~Scobel},
\author[MEPI]{R.~Shafigullin},
\author[Novosibirsk]{B.~Shwartz},
\author[Tuebingen]{T.~Skorodko},
\author[ITEP]{V.~Sopov},
\author[SINS]{J.~Stepaniak},
\author[ITEP]{V.~Tchernyshev\thanksref{d1}},
\author[SU]{P.-E.~Tegn\'er},
\author[UU]{P.~Th\"orngren Engblom},
\author[Dubna]{V.~Tikhomirov},
\author[WU]{A.~Turowiecki},
\author[Tuebingen]{G.J.~Wagner},
\author[UU]{M.~Wolke},
\author[Tsukuba]{A.~Yamamoto},
\author[SINS]{J.~Zabierowski},
\author[SU]{I.~Zartova},
\author[UU]{J.~Z{\l}oma\'nczuk}

\address[SU]{Stockholm University, Stockholm, Sweden}
\address[Tuebingen]{Physikalisches Institut der Universit\"at T\"ubingen, T\"ubingen, Germany}
\address[Dubna]{Joint Institute for Nuclear Research, Dubna, Russia}
\address[TSL]{The Svedberg Laboratory, Uppsala, Sweden}
\address[UU]{Uppsala University, Uppsala, Sweden}
\address[Hamburg]{Institut f\"ur Experimentalphysik der Universit\"at Hamburg, Hamburg, Germany}
\address[MEPI]{Moscow Engineering Physics Institute, Moscow, Russia}
\address[Juelich]{Institut f\"ur Kernphysik, Forschungszentrum J\"ulich, J\"ulich, Germany}
\address[Novosibirsk]{Budker Institute of Nuclear Physics, Novosibirsk, Russia}
\address[ITEP]{Institute of Theoretical and Experimental Physics, Moscow, Russia}
\address[SINS]{Soltan Institute for Nuclear Studies, Warsaw and Lodz, Poland}
\address[WU]{Institute of Experimental Physics, Warsaw, Poland}
\address[Tsukuba]{High Energy Accelerator Research Organization, Tsukuba, Japan}
\thanks[d1]{Deceased}

\begin{abstract}
The reaction  \pdHeh\ at threshold was used
to provide a clean source of $\eta$ mesons for decay studies
with the WASA detector at CELSIUS. The branching ratio of the
decay \hppee\ is measured to be (4.3$\pm$1.3$\pm$0.4)$\times$10$^{-4}$.
\end{abstract}
\begin{keyword}
eta meson decays, Dalitz decays
\PACS{13.20.-v,25.40.Ve,14.40.Aq}
\end{keyword}
\end{frontmatter}

\section{Introduction}
\label{sec:intr} \setcounter{section}{1}\setcounter{equation}{0}

Radiative processes involving one or two photons are responsible for
the most common decays of the lightest pseudoscalar mesons
$\pi^\circ$, \et\ and $\et'$.  They are accompanied, as understood
from Quantum Electrodynamics (QED), by a processes where a virtual
photon converts into an electron-positron pair.  The conversion decays
are suppressed by a factor of the order of the fine structure constant
$\alpha$.  The virtual photon probes the structure of the decaying
meson and the interaction region in the time-like region of
four-momentum transfer squared, $q^2$, which is equal to the invariant
mass squared of the lepton pair.  An extensive review of the
conversion decays of the light mesons is given by Landsberg
\cite{Land}.
  
Experimental information on the conversion decays of \et\ meson decays
is scarce. The most frequent \et\ decay into an $e^+e^-$ pair is
\heeg\ with a branching ratio ($BR$) of $(6.0\pm0.8)$\E{-3}
\cite{PDG06}.  Only a few hundred events were collected in recent
experiments \cite{Akm01,Ach01}. Another decay is \hppee\ which is
related to the radiative decay \hppg\ ($BR=(4.69\pm0.11)$\E{-2}
\cite{PDG06}).  The first observation of one candidate event for this
decay was reported from a hydrogen bubble chamber experiment by
Grossman, Price and Crawford \cite{Gro66} in 1966.  Recently
semileptonic decays of the \et\ mesons were studied by the CMD-2
Collaboration \cite{Akm01} using the radiative decay $\phi\to \eta
\gamma$ as the source of \et\ mesons and 4 event candidates of \hppee\
were identified with an estimated background of 0.5 event.  The
average value of the $BR$ extracted from the two experiments is
($4.0^{+5.3}_{-2.5})$\E{-4} \cite{PDG06}.

Predicted values for $BR(\hppee)$ are given in at least three papers.
Jarlskog and Pilkuhn \cite{Jar67} got an upper value of $3.1\times
10^{-4}$.  Faessler et al.\cite{Faessler:1999de} have calculated the
decay rate within the Vector Meson Dominance (VMD) model and obtained
a value ($3.6\pm0.6$)\E{-4}.  Picciotto and Richardson \cite{Pic93}
got a value ($3.2\pm0.3$)\E{-4} using a model which incorporates
vector mesons in the chiral perturbation theory Lagrangian.  In this
paper the influence of an intermediate $\rho$ meson was taken into
account both for pion and lepton pairs. The calculations are similar
to the ones for the $K_L\to \pi^+\pi^-e^+e^-$ decay
\cite{Majumdar:1970ak,Sehgal:1992wm}.

Further interest in the \hppee\ decay comes from tests of CP violation
in \et\ decays which are often motivated by corresponding tests in
$K_L$ decays. A recent prediction (from CP violation) and observations
of an asymmetry in the distribution of angles between the \pip\pim\
and the $\e^+\e^-$ production planes in $K_L\to \pi^+\pi^-e^+e^-$
\cite{Sehgal:1992wm,Alav,Lai} have triggered theoretical speculations
that a similar observation in \hppee\ decay might reveal unexpected
mechanisms of CP violation in flavor conserving processes.  It appears
that there could be (hypothetical) CP violating contributions which
are not constrained by the limits on the $\et\to \pi\pi$ $BR$ or the
measured limit of neutron electric dipole moment. A measurement done
with a sensitivity better than $10^{-2}$ for the asymmetry will
provide a stringent constraint for such mechanisms \cite{Gao02}.

\section{The experimental method}
\label{sec:exper}

In the experiment performed by the CELSIUS/WASA collaboration, the
\et\ mesons were produced in the \pdHeh\ reaction at 893~MeV incident
proton kinetic energy, very close to the threshold.  This reaction was
first employed as an \et\ source for decay experiments at the Saturne
II synchrotron at Saclay \cite{Ber88}.  The production of \et\ mesons
was tagged by measuring $^3$He at 0$^{\circ}$.  The \et\ production
cross section in the reaction increases to a plateau value of 0.4
$\mu$b at about 2~MeV excess energy, where the background from prompt
$pd\to ^3$He\pip\pim\ reaction is at the percent level \cite{May96}.
This tagging method enables to collect simultaneously an unbiased data
sample of all \et\ meson decays.

\vspace*{-5mm}
\begin{figure}[ht]
\begin{center}
\includegraphics[angle=270,width=\textwidth]{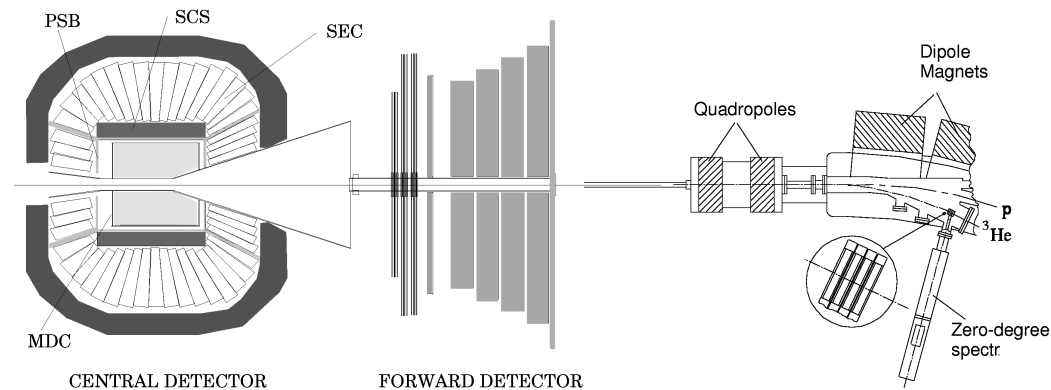}
\end{center}
\caption{%
Cross section of the WASA apparatus with Central Detector
(CD) built around the interaction region, Forward Detector (FD) and
Zero Degree spectrometer (ZD). The individual components are described
in the text.  }
\label{fig:1}
\end{figure}

\begin{figure}[ht]
\begin{center}
\includegraphics[width=0.44\textwidth]{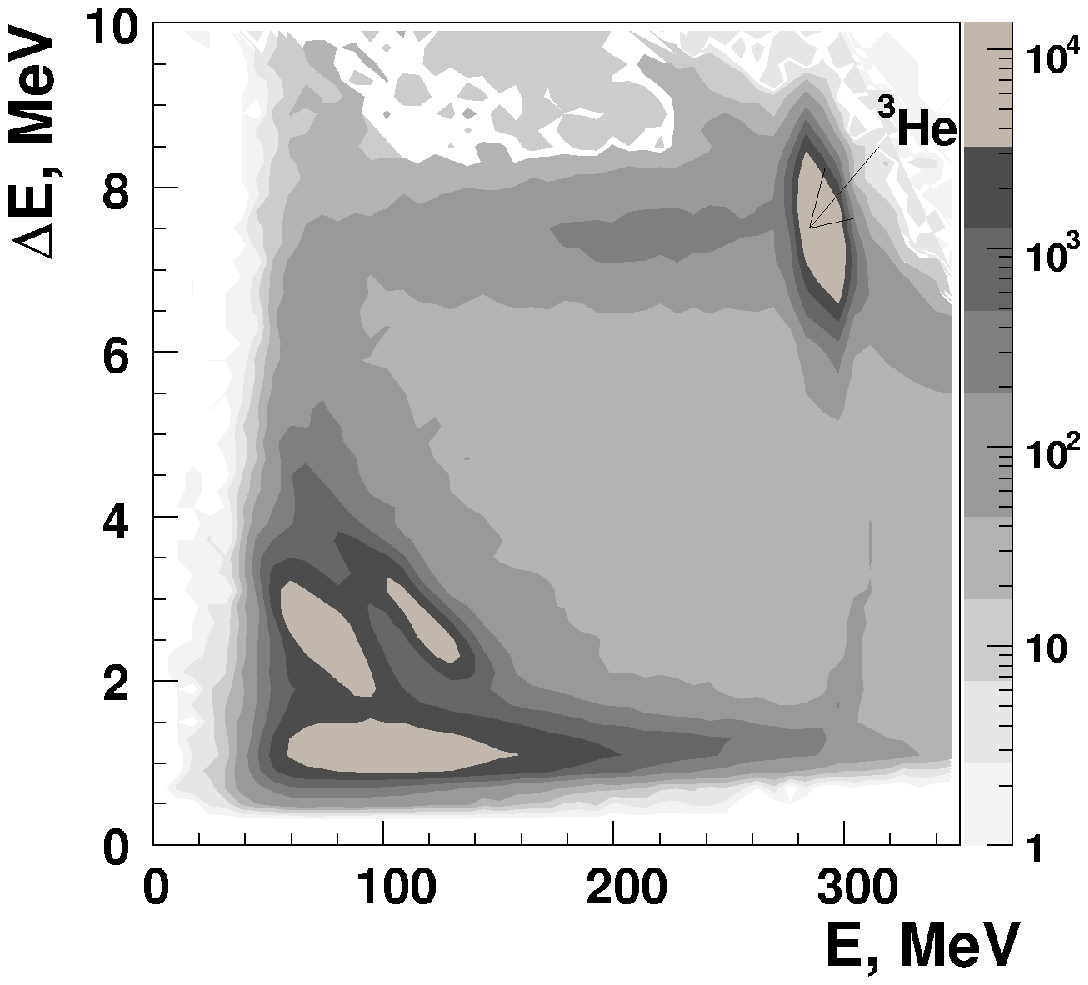}%
\hspace*{10mm}\includegraphics[width=0.38\textwidth,clip]{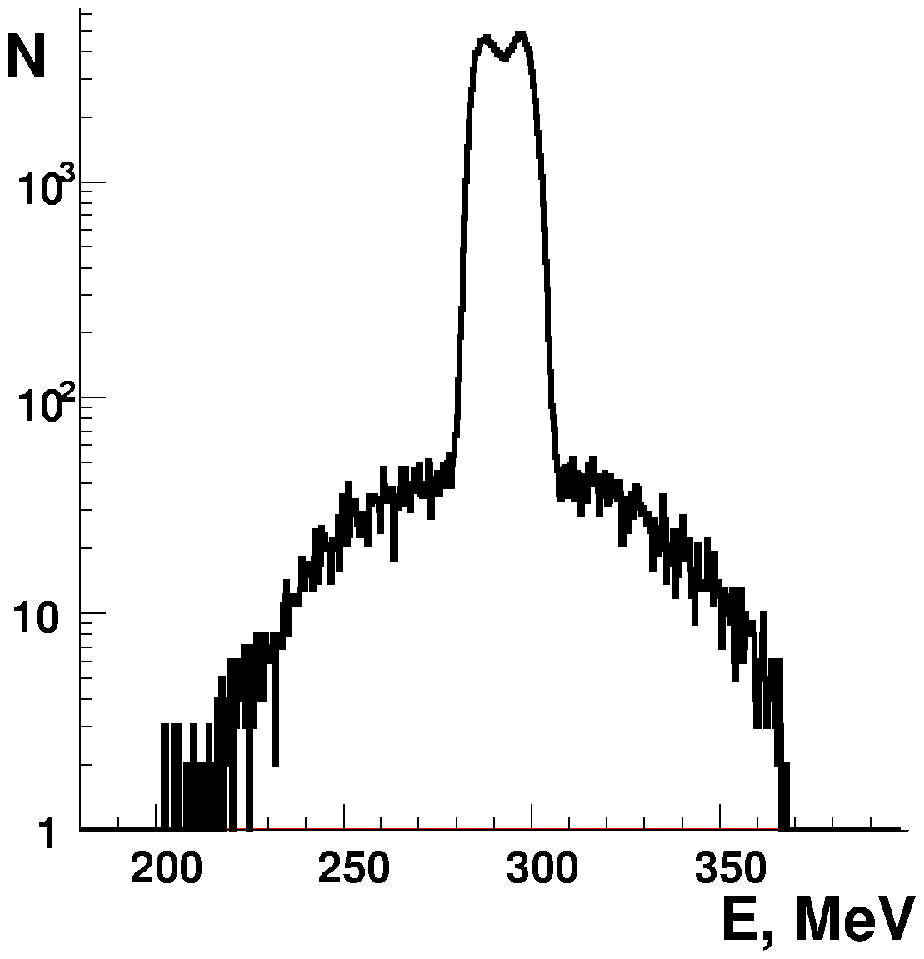}
\end{center}
\caption{%
(Left) $\Delta$E-E spectrum from the ZD, where $\Delta$E(E)
is the total energy deposition in the two silicon (germanium)
detectors indicating very clean $^3$He identification.  (Right) Energy
spectrum of $^3$He nuclei. The pronounced distribution around 300~MeV
corresponds to the \et\ production.  The cut-off at 200 and 380~MeV in
the distribution associated with 2$\pi$ production is a result of the
rigidity acceptance of the spectrometer.  }
\label{fig:2}
\end{figure}

In the CELSIUS/WASA experiment an internal deuterium pellet target was
used. The $^3$He nuclei recoiling at 0$^{\circ}$ were detected in a
zero-degree spectrometer (ZD) (Fig.~\ref{fig:1}). It used the
accelerator quadrupole and dipole magnets to focus and deflect the
$^3$He nuclei onto a tagging telescope, which was placed inside the
beam pipe 6.6 m downstream from the target \cite{Bar97}.  The
telescope, comprising two silicon detectors (total thickness 1.3 mm)
and two germanium detectors (total thickness 29.4 mm), was able to
stop $^3$He nuclei up to 400~MeV. Operated at liquid nitrogen
temperature, the telescope provided an energy resolution of 1.5~MeV
(FWHM) at 300~MeV \cite{Bar}.  The $\Delta$E-E spectrum in
Fig.~\ref{fig:2}(left), where $\Delta$E(E) is the total energy
deposition in the two silicon (germanium) detectors, shows that the
$^3$He nuclei are easily separated already in the raw data.
Fig.~\ref{fig:2}(right) shows the energy spectrum of $^3$He nuclei.
The pronounced distribution between 280 and 305~MeV is due to \et\
production.  The two peaks, indicated by the structure at the top of
the distribution, correspond to backward and forward emission of
$^3$He, respectively, in the center-of-mass frame.  The smoother
distribution is mainly due to direct 2$\pi$ production, yielding a
background of less than 2\% in the \et\ meson tagging.  The acceptance
for $^3$He nuclei from the \et\ production is approximately 50\% at
1~MeV above the threshold.

\vspace*{-2mm}

\begin{figure}[ht]
\begin{center}
\includegraphics[width=0.4\textwidth,clip]{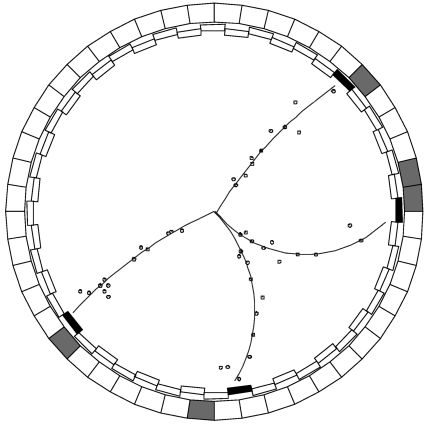}
\end{center}
\caption{%
Example of a candidate event for \hppee\ decay. Four charged tracks
are measured in the MDC, all particles leave the volume of the chamber
and deposit energy in the PSB and SEC (represented as the two outer
rings).  }
\label{fig:3}
\end{figure}

The \et\ decay products were detected in the central part of the WASA
detector (Fig.~\ref{fig:1}) consisting of an electromagnetic
calorimeter with over 1000 CsI(Na) crystals (SEC), a plastic
scintillator barrel detector (PSB) and a drift chamber (MDC) made of
thin--walled (25$\mu m$) aluminized mylar straw tubes placed in a 1
Tesla magnetic field produced by a very thin walled (0.18 $X_0$)
superconducting solenoid \cite{CW02} (SCS).  The design of the WASA
detector was optimized for measurements of decays with both electrons
and photons. To minimize the probability for photon conversion it
included a thin beryllium beam pipe with a wall thickness of 1.2 mm
(0.0034 radiation lengths only). Fig.~\ref{fig:3} shows an example of
a candidate event for \hppee decay registered in the CD.

\vspace*{-5mm}

\section{Analysis and results}
\label{sec:analysis}

The candidate events for the \et\ decay into $\pi^+\pi^- e^+ e^-$ were
selected by requiring four tracks caused by the charged particles in
the central drift chamber with a possible energy deposit in the SEC
and/or the PSB.  Two particles should carry positive and two negative
charge.  A candidate event could in addition have a calorimeter
hit-cluster not clearly associated with any of the tracks in the MDC
(such a cluster will be referred to later in the paper as a
\fakeg-cluster).  These events could not be discarded directly without
significant loss of acceptance (in the order of 35\%).  $MC$ studies
show that small but well separated additional clusters can arise from
the \hppee\ decay products due to secondary reactions in the
calorimeter material resulting in emission of neutral
particles. However, the energy deposit in such clusters is expected to
be small. Only events with an energy deposit below 80~MeV in the
\fakeg-clusters were accepted as candidates.

  Due to the clean tagging in \pdHeh, as discussed in the previous
section, the only major background for the \hppee\ decay can be due to
other \et\ decays into two charged pions.  A source of background
could be the \hppp\ decay with $\pi^0\to e^+e^-\gamma$ and the \hppg\
decay with the photon undergoing conversion in the detector
material. In order to reduce the latter background a vertex position
constraint ($\Delta$x,$\Delta$y =$\pm 5$~mm and $\Delta$z =$\pm
50$~mm) was used to ensure that all particles originated from the
interaction region.  That constraint removes more than 94\% of \hppg\
events with photon conversion.

The two-lepton invariant mass distribution
$d\Gamma/dq^2(\eta\to\pi^+\pi^-\e^+\e^-)$ can be expressed
approximately as $d\Gamma/dq^2=\Gamma(\eta\to\pi^+\pi^-\gamma)\left[
QED\right] |F(q^2)|^2$.  The QED term ($\left[ QED\right]$) is given
by \cite{Land}:

\begin{eqnarray}
\!\!\!\!\! [QED]&=&\frac{\alpha}{3\pi}
\sqrt{1\!-\frac{4m^2_e}{q^2}}\left[1\!+\frac{2m^2_e}{q^2}\right]\!\frac{1}{q^2}\!
\left[\left(1\!+\frac{q^2}{m^2_{\eta}-M^2_{\pi\pi}}\right)^2\!\! - \frac{4m^2_{\eta}q^2}
{(m^2_{\eta}-M^2_{\pi\pi})^2}\right]^{\frac{3}{2}}  \label{eq:one}
\nonumber 
\end{eqnarray}

where 
$m_e,m_{\eta}$ are the electron and \et\ masses and
$M_{\pi\pi}$ is the invariant mass of the two-pion system. 
The $q$ should be in the range $2m_e<q<m_{\eta}-2m_{\pi^\pm}$.
The QED term leads to a strong enhancement in
the $d\Gamma/d q^2$ distribution at the lowest possible $q$ values. 
The form-factor $F(q^2)$  modifies the
distribution mainly at large $q$. 
The  most common assumption for the  
form-factor  for \et\ and $\pi^o$ conversion decays  is  a  dipole
parametrization justified by $\rho^0$ dominance.

\begin{figure}[ht]
\begin{center}
\includegraphics[width=0.43\textwidth,clip]{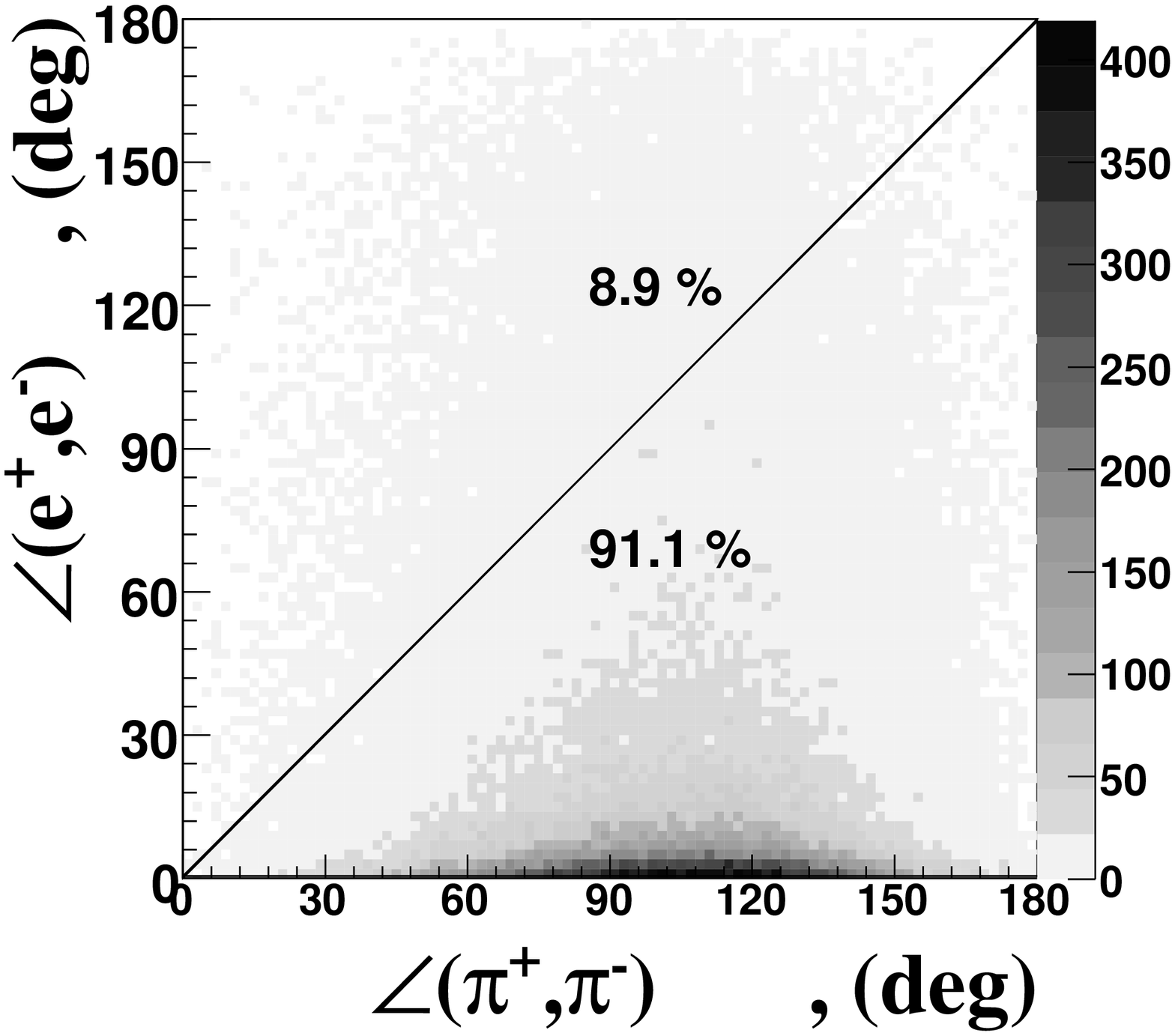}%
\hspace*{5mm}\includegraphics[width=0.43\textwidth,clip]{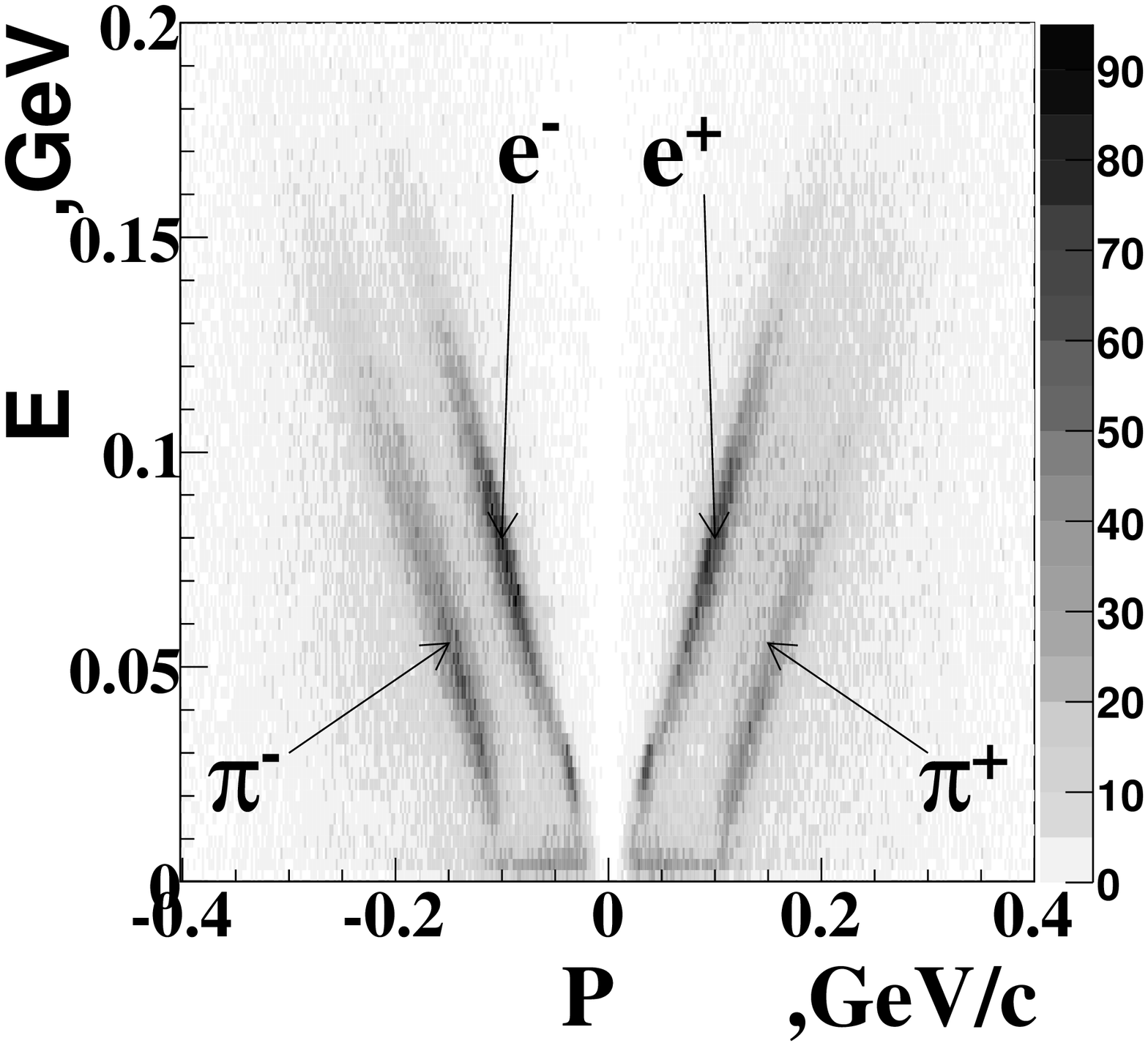}
\end{center}
\caption{%
Particle identification for the $MC$ sample of \hppee\ decay.  (Left)
Angle between the two pions (X-axis) versus the angle between the two
leptons (Y-axis). The latter angle is smaller in 90\% of cases.
(Right) E-P distributions for all tracks with energy released in the
SEC versus reconstructed particle momentum.  The bands due to pions
and electrons are well separated.}
\label{fig:4}
\end{figure}

The invariant mass is closely correlated to the opening angle between
the leptons leading to a sharp peak at small opening angles.  This
feature of the process is used for particle identification.  In order
to distinguish between pions and electrons, the opening angle between
two particles with opposite charges is calculated. The pair with the
smallest opening angle is most likely the positron-electron pair.  The
efficiency of such a separation is demonstrated in
Fig.~\ref{fig:4}~(left) using Monte Carlo ($MC$) simulation with the
\hppee\ mechanism given by the above formula. In addition an
energy-momentum (E-P) method of particle identification is applied for
all the particles which did not stop in the MDC.  The energy deposit
(dE) in the SEC and/or the PSB (E = dE$_{SEC}$ + dE$_{PSB}$) is
combined with the momentum (P) calculated from the track curvature
(the sign gives the charge of the particle).  Fig.~\ref{fig:4} (right)
shows the simulated E-P distributions for all reconstructed tracks
with associated energy deposition in the SEC.  In that case the pion
and electron bands, which are well separated, can be used to check and
improve the identification based on the opening angle.  The overall
identification is correct in 90\% of the cases for all tracks in the
$MC$ event.

\vspace*{-2mm}
 
\begin{figure}[ht]
\begin{center}
 \includegraphics[width=0.5\textwidth,clip]{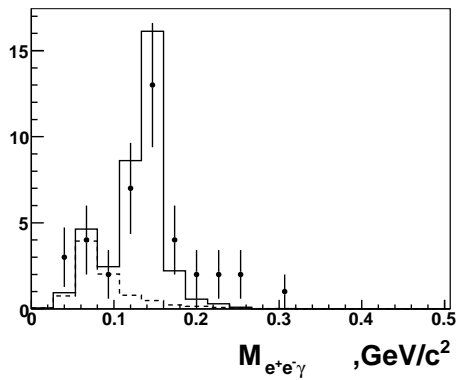}
\end{center}
\caption{%
The experimental invariant mass of \eefakeg\ for the events with the
\fakeg-cluster. The pronounced $\pi^0$\ peak in the experimental mass
spectrum (black points) indicates a significant admixture of the
background events. Superimposed are: a sum of the appropriately
weighted $MC$ simulation of \hppee, \hppp\ and \hppg\ decays (full
line) and \hppee\ decay (dashed line).}
\label{fig:5}
\end{figure}

In order to suppress the background from the \hppp\ decays two
constraints optimized by the $MC$ studies were used.  The missing mass
of the pions and leptons was constrained around the $^3$He mass (in
the range from 2.6~GeV/c$^2$ to 2.85~GeV/c$^2$).  This reduces the
number of \hppee\ and \hppp\ events by 10\% and 47\% repectively
according to the $MC$ studies.  Some of the remaining events had
so-called \fakeg-clusters in the calorimeter present (hit-clusters
unassociated with a track in the MDC).  Fig.~\ref{fig:5} presents the
invariant mass of the \eefakeg\ system for the experimental sample
with four charged particles and a neutral cluster in the final
state. We find a clear peak around $\pi^0$ mass in accordance with the
simulation of the \hppp\ decay (with $\pi^0\to e^+e^-\gamma$).  Thus,
for the class of events with a \fakeg-cluster, the invariant mass of
the \eefakeg\ system was calculated and an event was identified as a
background event, if the mass was greater than 120~MeV/c$^2$.

\begin{figure}[ht]
\begin{center}
 \includegraphics[width=0.5\textwidth]{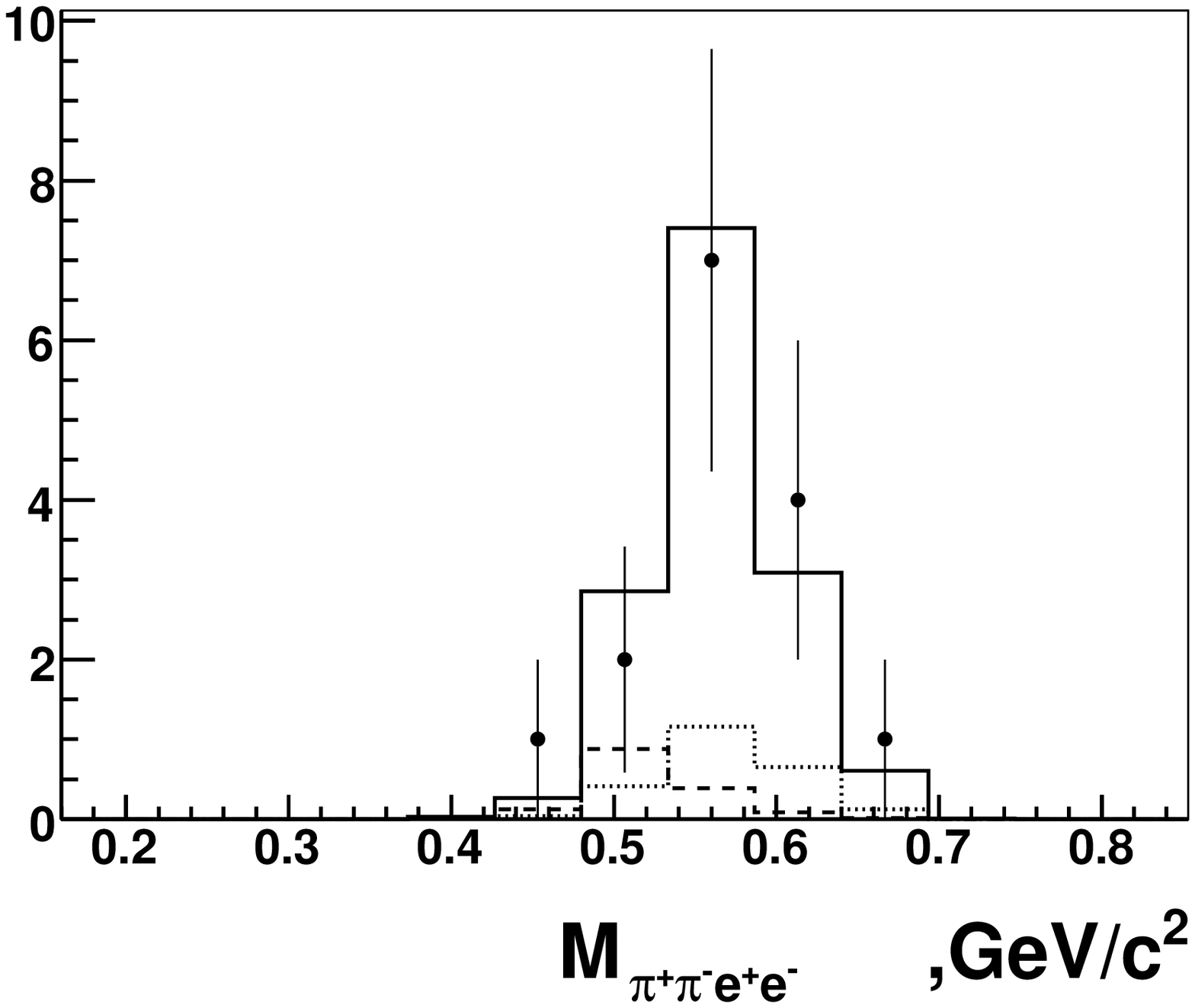}%
 \includegraphics[width=0.5\textwidth]{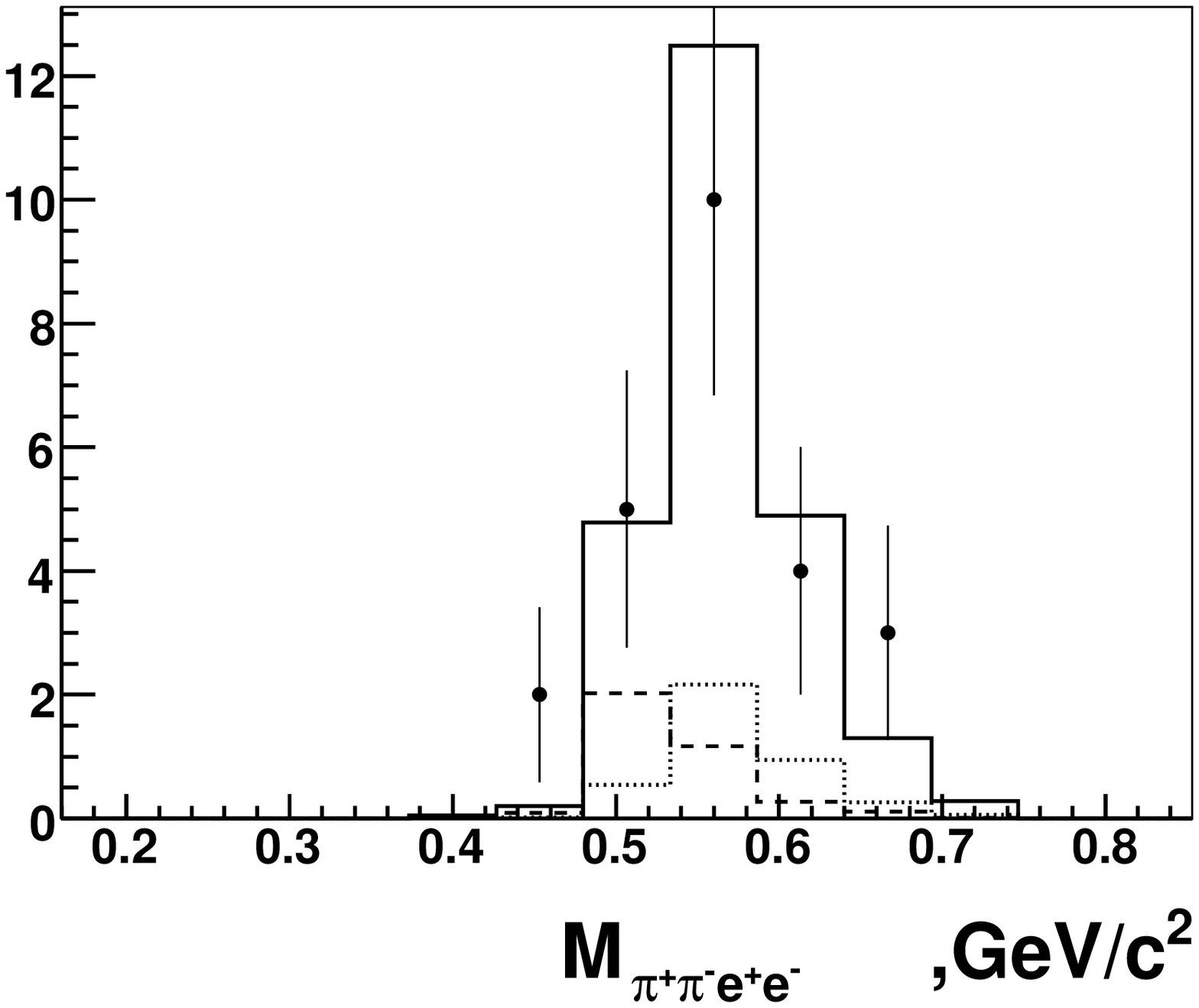}
\end{center}
\caption{%
 The invariant mass distribution, M$_{\pi^+\pi^-e^+e^-}$, plotted for
events without the \fakeg-clusters unassociated with the MDC tracks
(left) and for all events (right) (black points in both figures).  The
thick line represents the $MC$ distribution corresponding to the sum
of the signal (\hppee) and the background decays. In addition the
contributions of the background from \hppp\ and \hppg\ decays are
given by the dashed and dotted line respectively.}
\label{fig:6}
\end{figure}

 The invariant mass distribution, M$_{\pi^+\pi^-e^+e^-}$, is plotted
in Fig.~\ref{fig:6} for the final data sample, first only for events
without the \fakeg-clusters (left) and later using all events (right).
The experimental distribution is compared in both cases to the $MC$
prediction taking into account the dominant background channels
(\hppp\ and \hppg) and the shape of the signal distribution.  The
contribution of the background events to the peak is estimated to be
4.6 and 7.7 events (in the left and right plot respectively) by taking
into account the known $BR$ and acceptance from the $MC$ studies.
(The background contributions are represented in the figures by the
broken lines).  The remaining events in the plots are attributed to
the \hppee\ decay.  This gives a number of reconstructed \hppee\
events for the right plot in Fig.~\ref{fig:6} equal to
16.3$\pm$5.0$_{stat}\pm$2.0$_{syst}$ (the systematic error includes
the ambiguity of the background contribution). The overall detection
efficiency for \hppee\ was estimated to be 16.5$\pm$0.2\% where the
error is dominated by the uncertainty in acceptance calculation due to
extrapolation using different reaction models.

The $BR$ of \hppee\ was obtained by normalizing to a sample of \hppp\
decays ($BR=(2.27\pm0.04)$\E{-1}\cite{PDG06}) and \heeg\ decays
($BR=(6.0\pm0.4)$\E{-3}\cite{PDG06}) collected simultaneously.  The
data were analyzed using the same track finding processing and
compatible cuts.  This assures that many sources of errors, like
uncertainties in particle identification and track finding efficiency,
will be reduced or cancel.

 The selection criteria for the process \hppp\ include requirement of
two tracks with opposite charges identified as pions by the E-P
method, two neutral clusters with energy E$>$\ 50~MeV each and an
invariant mass of the two photons to be within the range 90 to
170~MeV/c$^2$.  Systematical uncertainties of the normalization were
estimated by varying the constraints during sample selection.  This
lead to a change of the normalization factor within a 5\% interval,
what was included in the systematic error.  The number of the
identified events for this process was 17100$\pm$130 and the
acceptance was estimated to be (33.4$\pm$1.9)\% (the error is a
cumulative systematic error of the normalization and includes the $MC$
model dependence, the background and the sample selection).

A cross-check of the normalization with the \heeg\ decay was done to
test the $MC$ predictions for the electron reconstruction efficiency.
In the collected sample a clear peak of 270 events of the \heeg\ decay
has been observed after application of cuts on the charge balance, the
overall missing mass and the invariant mass of the two leptons.
Taking into account the calculated acceptance (20\%) and assuming
$BR=(6.0\pm0.8)\times10^{-3}$ \cite{PDG06} for the decay we got a
normalization factor smaller by 5\% from the one obtained with \hppp\
decay, which is within the error of the given $BR$.  The variation of
cuts on the $e^+e^-\gamma$ missing mass and/or replacing the cut on
the lepton pair invariant mass by the cut on their relative angle lead
to a variation of the normalization factor within 8\% which is within
one standard deviation of the number of \heeg\ events.

Taking the obtained normalization factors and their statistical 
and systematical uncertainties the result for the $BR$ of the
\hppee\ decay is (4.3$\pm$1.3$\pm$0.4)$\times$10$^{-4}$ based on 
16.3$\pm$4.9$_{stat}\pm$2.0$_{syst}$ identified events.

 It has been demonstrated that with the applied data selection the
background channels are well understood. Our result increases notably
 the experimental information available for this \et\ decay channel.
 The search for CP violation mechanism of \cite{Gao02}, however,
 clearly requires much more data than what has been collected so far.

\begin{ack}
We are grateful to the personnel at The Svedberg Laboratory for their
support during the course of the experiment.  This work was supported
by the European Community under the ``Structuring the European
Research Area'' Specific Programme Research Infrastructures Action
(Hadron Physics, contact number RII3-cT-204-506078) and by the Swedish
Research Council.

\end{ack}


\begin{thebibliography}{99}
\bibitem{Land} L.G.~Landsberg, Phys. Rep. 128 (1985) 301.
\bibitem{PDG06} W.-M.~Yao et al. [PDG], J. Phys.  G 33 (2006) 1.
\bibitem{Akm01} CMD-2 Collab., R.R.~Akhmetshin et al., Phys. Lett. B 501 (2001) 191.
\bibitem{Ach01} M.N.~Achasov et al., Phys. Lett. B 504 (2001) 275. 
\bibitem{Gro66} R.A.~Grossman, L.R.~Price and F.S.~Crawford, Phys. Rev. 146 (1966) 993.
\bibitem{Jar67} C.~Jarlskog and H.~Pilkuhn, Nucl. Phys. B 1 (1967) 264.
\bibitem{Faessler:1999de} A.~Faessler, C.~Fuchs and M.~I.~Krivoruchenko, Phys. Rev. C 61 (2000) 035206.
\bibitem{Pic93} C.~Picciotto and S.~Richardson, Phys. Rev. D 48 (1993) 3395.
\bibitem{Majumdar:1970ak} D.~P.~Majumdar and J.~Smith,
  Phys. Rev. 187 (1969) 2039.
\bibitem{Sehgal:1992wm} L.~M.~Sehgal and M.~Wanninger,
  Phys. Rev. D 46 (1992) 1035
  [Erratum-ibid. D 46 (1992) 5209].
\bibitem{Alav} KTeV Collab., A.~Alavi-Harati et al., Phys. Rev. Lett. 84 (2000) 408.
\bibitem{Lai} NA48 Collab., A.~Lai et al., Eur. Phys. J. C 30 (2003) 33. 
\bibitem{Gao02} Dao-Neng Gao, Mod. Phys. Lett. A 17 (2002) 1583.
\bibitem{Ber88} J.~Berger et al., Phys. Rev. Lett. 61 (1988) 919.
\bibitem{May96} B.~Mayer et al., Phys. Rev. C 53 (1996) 2068.
\bibitem{Bar97} Chr.~Bargholtz et al., Nucl. Instrum. Meth. A 390 (1997) 160.
\bibitem{Bar} Chr.~Bargholtz et al., Instrum. Exp. Tech. 49 (2006) 461.
\bibitem{CW02} CELSIUS/WASA Collab., J.~Zabierowski et al., Phys. Scr. T 99 (2002) 159.

\end{thebibliography}
\end{document}